\documentstyle[prl,aps,twocolumn]{revtex}

\begin{document}

\title{Nanomechanical vibrating wire resonator for phonon spectroscopy
in Helium}
\draft
\author{Andreas Kraus, Artur Erbe, and Robert H. Blick}
\address{
         Center for NanoScience and Sektion Physik,
Ludwig-Maximilians-Universit\"at, Geschwister-Scholl-Platz 1,
80539 M\"unchen, Germany. \\
        }
\date{\today}
\maketitle

\begin{abstract}
We demonstrate how to build a vibrating wire resonator for phonon
excitation in liquid helium. The resonator
is designed as a nanoscopic mechanically flexible beam machined out of
a semiconductor/metal-hybrid. Quenching of the mechanical resonance
around 100~MHz by phonon excitation in liquid $^4$He at 4.2~K is
shown. First measurements operating the nano-resonator
in a dilution of $^3$He/$^4$He at 30~mK are presented.
\pacs{62.30.+d, 67.40.Vs, 07.10.Cm}
\end{abstract}
\newpage

\vspace*{0.8cm}

Vibrating wire resonators (VWR) are standard bolometers for the quantum fluids
$^3$He/$^4$He~\cite{lancaster}. The
general idea is to immerse a metallic wire into
superfluid helium and to induce a mechanical vibration by applying a
magnetic field while 
simultaneously sending an alternating current through the wire. Usually these
wires are resonating
at several kHz with deflections on the order of some microns. The
wire moving in the
superfluid will generate a phonon flux, which is directed parallel to
the plane of motion of the wire. With these macroscopic wires excitations 
in $^4$He are usually not found, since the Landau critical velocity is not
in the
accessible range.

Here we present a new approach to build such phonon radiators for
applications
in spectroscopy of quantum liquids. In contrast to previous resonators
operating in the kHz-range only, we focus on the realization of wires or more
specifically on suspended hybrid Si/metal-beams 
with nanometer dimensions and hence resonance frequencies up to
1~GHz~\cite{cleland,nonlin}. We will first discuss
processing of the nanostructures and then proceed to the experimental
setup. Moreover, we show measurements allowing us to calibrate the
attenuation of
the nano-VWR in liquid $^4$He. Finally, we show first results on
resonating nano-VWRs in $^3$He/$^4$He at 30~mK.

The beams are machined from commercially available Silicon-on-Insulator
(SOI) wafers
with a top layer thickness of 205~nm and a sacrificial layer of 400~nm.
Optical lithography and evaporation of 180~nm NiCr-Au/Ti are used to define
the necessary bond pads and an etch mask (Ti), which is then removed in the
wet
etch step. An electron beam writer (JEOL 6400)
is used to define the nanomechanical resonator, consisting finally
of the metallic line on top (Au) and the Si supporting membrane.
In the following step the sample is
etched in a reactive ion etcher (RIE) using CF$_4$  as an etchant. 
The section of the sample, which is not covered by metal, is milled
down by 200 nm. Finally the sacrificial layer is removed using
diluted~(2\%) hydro-fluoric acid (HF) at an etch-rate of 10~nm/sec.

In the right inset of Fig.~1 the suspended beam is shown in a scanning
electron beam micrograph:
The beam is freely suspended between two tuning gates, coupling
capacitively. The beam has a length of 1~$\mu$m, a  width of  200~nm, and the
gates are covered by a 50~nm Au layer.
The gates can be applied to tune the mechanical resonance by
biasing up to about 10~V. 
For the measurements we employ a Hewlett-Packard network analyzer (HP
8751A), monitoring amplitude and phase of the resonator simultaneously with
Hz-resolution. The beam's resonance is excited by the alternating current
at radio
frequencies and the static magnetic field applied in plane. At resonance
a magnetomotive force is induced, which in turn
can be read out. In our case, the reflection of the signal is
determined, which is transduced upon the beam. 
For sensitive detection of the induced voltage signal a double sided
high-gain (15~dB) amplifier is brought into the line feeding the
circuit. The amplifier is optimized for a frequency response between 1 and
500~MHz. The resistance of the cables and the leads in the sample
holder is 50~$\Omega$, the dc-resistance of the sample is found to be
30~$\Omega$, thus a fairly good impedance matching is guaranteed.

First we measured resonance curves characterizing the elastic
properties of the beam in linear (L)
and nonlinear (NL) response; these measurements are shown in Fig. 1. The
resonance frequency is around 96~MHz at 4.2 K and shifts to lower values
when the resonator enters the nonlinear regime. 
The reflection factor is defined by

\begin{displaymath}
    r = \frac{P_{in}}{P_{out}},
\end{displaymath}

obviously it decreases with increasing power. The
resonator can be modelled 
as a typical Duffing oscillator using:

\begin{equation} \label{duffing}
    y''(t) + \gamma y'(t) +\omega_{0}^2 y(t) + k_{3}y^3(t) = A \sin
(\omega
    t),
\end{equation}

where $y$ is the elongation, $\omega_0 = 2\pi f_0$ the eigenfrequency 
of the beam,
$\gamma = \omega_0 / Q = 4.5 \times 10^{5}$~s$^{-1}$ 
represents the damping constant and the driving amplitude is
given by $A = 1.4 \times 10^{5}$~m/s$^2$ at $-63$~dBm and 
$2.8 \times 10^{6}$~m/s$^2$ at $-37$~dBm.
The shift to lower frequencies implies that for the cubic constant
in the Duffing equation we have $k_{3} < 0$. Attenuation commonly
shifts the eigenfrequency of a linear resonator to lower frequencies
and deteriors the
quality factor. In the measured resonator a quality factor of 2000
was observed. In a nonlinear resonator the eigenfrequency changes in
addition with
increasing attenuation, since the amplitude is reduced and the
resonator is driven closer to the linear regime. In our measurement
this shift leads to higher frequencies. Thus, if the
resonator is damped by liquid helium, a shift in the oscillating
frequency first to higher and then to lower frequencies
and a broadening of the resonance is expected.

Fig.~2(a) shows the measurement of the reflected power during 
filling of the sample holder with $^4$He when the wire is driven
in linear (L) response. The sample holder was cooled to 4.2~K.
The attenuation by the gas leads to a decrease of  the 
resonance amplitude, a frequency shift to lower frequencies and a
broadening of the resonance. When liquefication of the $^4$He starts (shown
in the last trace), the resonance disappears completely.
As seen in this figure we find even in the linear regime of the
resonator's response a dispersion when the $^4$He content is increased. This
corresponds to an effective momentum transfer to the $^4$He-atoms impinging
onto the resonator. We estimate this momentum transfer per atom to be
$1.9 \times 10^{-44}$~Ns at $p \cong 20$~mbar, $T = 4.2$~K, and at an 
excitation power of the network analyzer of $-68$~dBm. 

The plane of phonon emission can be chosen according to the direction
of the magnetic field, as sketched in the inset of Fig.~2(a). In the present
case the magnetic field was applied perpendicular to the plane of
the sample, hence excitations are propagating
in-plane away from the beam. Obviously the electrodes in the current
setup disturb the free propagation of phonons. This might cause a
back-flow and hence turbulences in the heat flow. 
Repeating the measurement in the nonlinear (NL) regime enables us to observe
the motion of the beam even in the liquid phase. This measurement is
depicted in Fig.~2(b): The first lower  
six curves are taken in $^4$He gas while the upper four curves
are taken in the liquid phase.
As seen the traces are changing drastically in the gaseous phase at some
10~mbar,
since with raising pressure the attenuation of the beam's motion is
increased due to the enhanced scattering of $^4$He-atoms. On the other
hand damping in the liquid depends only on the viscosity, which
does not change when supplying more helium. Thus the curves change
only slightly when the beam is immersed in the liquid phase. Hence, we 
can conclude that it is possible to create excitations in liquid $^4$He.

The main focus naturally is the possibility to apply the nanomechanical
resonators as phonon generators in superfluid helium. This is especially 
interesting, since the motion of the beam is fast compared to the critical
velocity in $^4$He. 
Until the beam reaches the
velocity, which is required to create vortex states
no attenuation should be present, since
the fluid is not excited by the motion of the beam~\cite{landau1}.
Further acceleration leads to the excitation of vortex states in the
fluid resulting in an increased energy consumption.
This can be seen in a flattening of the resonance curves. In $^3$He the
critical velocity is given by the energy required to break the superfluid 
state, which is in
the mm/s range \cite{lancaster}. In $^4$He it is given by the creation of
vortex states at
low pressures and the excitation of rotons at higher pressures. The
corresponding velocities are about 25~m/s. With
cold neutron scattering experiments~\cite{henshaw}
even the zero pressure critical velocity has been measured ($v_{c} =
60~{\rm m/s}$).
Therefore vibrating wire experiments have only been
successful in $^3$He, while in $^4$He the critical velocity has 
been reached with accelerated charged ions only~\cite{rayfield}.

In order to verify that we are able to reach 
the critical velocities, we monitored the velocities of the beam
depending on the input power and the amount of $^4$He in the sample
holder. We did this by calculating the maximum amplitude $Y_{0,max}$
during one cycle of the oscillation using the equation:

\begin{equation}
    Y_{0,max}=\frac{lBI_{0}}{2m_{eff}\omega_{0} \mu}.
\end{equation}

Here $\mu = 2.23 \times 10^{5}$~s$^{-1}$ is the damping coefficient of the 
mechanical system, $l = 2 L/ \pi = 1.165$~$\mu$m and $m_{eff} = 3.418 \times 10^{-16}$~kg 
are the effective length and effective mass of the resonator, $B$ the
magnetic field.
The amplitude of the input current at $-63$~dBm is $I_{0} = 4.09 \times 10^{-6}$~A 
and at $- 37$~dBm it is $8.16 \times 10^{-5}$~A. Subsequently we obtain
$Y_{0,max} = 9.79 $~nm. The parameters of the Duffing eq.~(\ref{duffing}) can be extracted
from the measurements in the nonlinear regime. The velocity is then
estimated via

\begin{displaymath}
    v \propto fY_{0,max}
\end{displaymath}

In the linear regime we find a linear dependence on the input
power as long as no $^4$He is added. The maximum velocity in this case
is 20~m/s. When filling with $^4$He, this velocity is decreased until
it finally reaches zero in the case of liquid helium (see Fig. 2).
In the nonlinear regime we begin at a velocity of 20~m/s, which is
then reduced to 5~m/s in the liquid. The decrease of the velocity is
mainly due to the deterioration of the quality factor.
To reach the critical velocities in $^4$He we have to increase the
velocity by a factor of 5, but this should be sufficient to observe
excitations of the superfluid~\cite{zwerger99}. 

For the observation of such excitations the sample is mounted in the
mixing chamber of a dilution refrigerator. The suspended beam is 
immersed in a dilution of 6\% $^3$He in superfluid $^4$He and cooled
to 30~mK. The resulting measurements are depicted in Fig.~3: 
As seen the shapes of the resonances strongly varies from the original
one determined at 4.2~K. Each resonance amplitude is monitored at different
magnetic field values, hence avoiding spurious effects of the sample holder.
We assume that the peculiar resonance shape  is caused by the increased
attenuation in the superfluid mixture. This seems to be counterintuitive, 
since conventional tuning fork resonators (TFRs) show less attenuation in 
superfluid phase of $^4$He~\cite{karrai}. These macroscopic resonators are
operated at some 10~kHz with typical dimensions in the mm-range, compared
to 100~MHz in our case. 
Moreover, a
similar strong attenuation is observed when surface acoustic wave
transducers
are immersed into superfluid $^4$He at 1~K, where the attenuation is enhanced
strongly~\cite{wixforth}, although the opposite behavior is expected. 
Hence, we can conclude that this strong attenuation is caused by the 
excitation of phonons and even rotons in the superfluid $^4$He (here we
assume the $^3$He content to be negligible).
An additional boundary resistance
like the Kapitza resistance $R_K \sim 1/T^3$~\cite{kapitza} might be ruled
out, since it should be observed for the TFRs as well. 

In the measurements shown in Fig.~3 the resonance frequency drops
continuously from 95.8~MHz (a), to 93.6~MHz (b), and to 72.2~MHz (c).
This we attribute to the gradual disintegration of the Si supporting 
structure, while the metal on top remains flexible until it is melted. Since
the Si beam is stiffer than the top metal a gradual decrease of the resonance
frequency is expected, when the Si beam is collapsing.  
A certain drawback of this resonator is the non-superconducting metal on
top of the beam. A crude estimation of the Joule heating dissipated by the
wire yields
an energy of $\cong 1.10 \times 10^{-20}$~J per cycle, while the energy
for excitations in superfluid $^4$He
is on the order of $3.04 \times 10^{-23}$~J. 
Hence, we observe mostly black body radiation and not
ballistic phonon emission. This problem can be overcome by using a
superconducting metal on top of the suspended beam. 
However, from the data we obtained so far we can conclude that it is
possible to create excitations in liquid as well as superfluid $^4$He. Although,
the phonon emission is not directed but isotropic, we are confident that 
a superconducting wire vibrating without dissipating the electromagnetic
fields will create directed acoustical phonons propagating in superfluid $^4$He.
Furthermore, since a frequency range of 1~GHz seems to be a realistic goal with current
lithography, velocities
up to ten times larger than what we achieved are possible to reach. Thus
even the regime of roton
excitations might be accessible.
Another experiment now in reach is the determination of the
dispersion relation of phonons in $^4$He.
For the low energetic excitations the dispersion 
is purely acoustical up to frequencies of some 100~GHz. It is
therefore possible to determine the dispersion of these
excitations in the low energy limit ranging from some MHz up to
possibly 1~GHz. Moreover, the eigenfrequency of the nanoresonators can
easily be varied by a tuning gate.

In summary we presented measurements on nanomechanical
resonators in gaseous and liquid $^4$He and in $^3$He/$^4$He at 30~mK. 
These measurements demonstrate strong attenuation of 
the mechanical excitations of the resonator.  
Since the elongation can be
tuned even in the liquid, the resonators can be applied as nano-VWR
in
order to create vortex state and roton excitations in superfluid
$^4$He and $^3$He.
In recent experiments quantum effects in superfluid $^3$He have
been observed \cite{pereverzev}. In order to observe quantum
effects the objects moving in
the liquid have to be smaller than the superfluid healing length.
This length, which determines the width of the vortex states as
well~\cite{tilley,packard},
is about 50~nm in superfluid $^3$He. Silicon beams
with a width of 80~nm have already been fabricated \cite{laura}. Thus
quantum effects in $^3$He are in the accessible range for those
nano-VWRs. Our calculations show that the velocities required for these excitations
are well in the experimentally accessible range.
Although, a clear signature
of directed phonon or roton excitation was not yet found, we presume that
by replacing the non-superconducting metal on top of the electromechanical
resonator by a superconducting one, we will obtain new insight into 
the excitation mechanisms of quantum fluids on a nanoscopic scale. 

We like to thank J.P. Kotthaus for his support
and W. Zwerger, and P. Leiderer for discussion.
Also we like to thank S. Manus for technical help. This work 
was funded in part by the Deutsche Forschungsgemeinschaft (DFG). \\


\newpage

\figure{Fig.~1:
Fundamental resonance of the Si/Au-beam: Shown is the reflection
coefficient of the
beam from the linear (L) into the nonlinear (NL) regime at 4.2~K and 20~mbar
He-gas pressure.
Left inset: Circuit diagram of the experimental setup. The reflected
signal is amplified and detected by the network analyzer.
Inset on the lower right: Scanning electron beam micrograph of the
vibrating
wire resonator. The suspended beam consists out of a silicon supporting
structure and a metallized top layer. The gates close by can be applied
for electrostatically tuning the beam's mechanical resonance.
        }
\label{one}

\figure{Fig.~2:
(a) Attenuation of the mechanical resonance amplitude by increasing $^4$He
pressure as indicated until liquefication in the linear (L) regime (-63~dBm
input power, $T = 4.2$~K). The maximum velocity of the beam $v_{max}$ has been
calculated for each curve and is given on the left hand side. 
In the inset a scheme for directed phonon generation is sketched. The
magnetic field is oriented perpendicular to the high frequency current
$I$. (b) Resonance curves of the beam in the nonlinear (NL) regime 
at different $^4$He fillings (4.2 K). Again the maximum velocity of the beam
has been calculated. The motion is not suppressed completely in 
liquid $^4$He, thus allowing vibrational excitations.
       }
\label{two}

\figure{Fig.~3:
Resonator oscillating in a dilution of $^3$He/$^4$He at 30~mK. The different
traces (a,b,c) are taken with the same device by increasing the magnetic field
each time. The continuous decrease of the eigenfrequencies from (a) to (b)
and (c) is caused by the disintegration of the Si supporting structure of the
suspended beam (for details see text). 
     }
\label{three}

\end{document}